\documentclass[twocolumn,dvipsnames,usenames]{aastex631}
\usepackage[utf8]{inputenc}
\usepackage{enumitem,graphicx}
\usepackage{xspace} 
\usepackage{booktabs}
\usepackage{upgreek} 
\usepackage{acronym}

\newcommand{\LIGOlabMIT}{\affiliation{MIT LIGO Laboratory, Massachusetts Institute of Technology, Cambridge, MA 02139, USA}}
\newcommand{\MKI}{\affiliation{MIT Kavli Institute for Astrophysics and Space Research, Cambridge, MA 02139, USA}}

\acrodef{bbh}[BBH]{binary black hole}
\acrodef{nsbh}[NSBH]{neutron star--black hole}
\acrodef{bns}[BNS]{binary neutron star}
\acrodef{lvk}[LVK]{LIGO-Virgo-KAGRA}
\acrodef{em}[EM]{electromagnetic}
\acrodef{gw}[GW]{gravitational wave}
\acrodef{agn}[AGN]{active galactic nuclei}
\acrodef{snr}[S/N]{signal-to-noise ratio}
\acrodef{dtd}[DTD]{delay time distribution}
\acrodef{sfr}[SFR]{star formation rate}
\acrodef{cbc}[CBC]{compact binary coalescence}
\acrodef{grb}[GRB]{gamma ray burst}
\acrodef{xg}[XG]{next generation}
\acrodef{ce}[CE]{Cosmic Explorer}
\acrodef{et}[ET]{Einstein Telescope}

\date{\today}

\begin{document}

\title{Identifying the hosts of binary black hole and neutron star-black hole mergers \\with next-generation gravitational-wave detectors} 

\correspondingauthor{Geoffrey Mo}
 \email{gmo@mit.edu}
\author[0000-0001-6331-112X]{Geoffrey Mo} \MKI \LIGOlabMIT 
\author[0000-0002-7778-3117]{Hsin-Yu Chen} \MKI \LIGOlabMIT
\affiliation{Department of Physics, University of Texas, Austin, TX 78712, USA}

\begin{abstract}
The \acl{lvk} Collaboration has detected over one hundred compact binary mergers in \aclp{gw}, but the formation history of these binaries remains an open question. 
Finding the host galaxies of these mergers will provide critical information that reveals how these binaries were formed. 
However, without an \acl{em} counterpart, localizing \acl{gw} events to their hosts is challenging with the current generation of \acl{gw} detectors. 
Next-generation detectors will localize some compact binary mergers to a small volume that allows for direct association with their hosts. 
To demonstrate the promise these detectors hold, we simulate a population of \acl{bbh} and \acl{nsbh} mergers using a next-generation \acl{gw} network comprised of \acl{ce} and \acl{et}. 
We find that $\sim 4 \%$ of \acl{bbh} events within a redshift of 0.5 and $\sim 3\%$ of \acl{nsbh} events within a redshift of 0.3 will be localized to a volume smaller than $100$ Mpc$^3$, the volume in which we expect only one likely host galaxy.
With the astrophysical merger rate estimated from the \acl{lvk} Collaboration's third observing run, we expect to precisely localize one \acl{bbh} event every eight days and one \acl{nsbh} event every 1.5 months.
With three years of \acl{gw} observations ($\mathcal{O}(100)$ \acl{bbh} mergers with host associations), we will be able to distinguish whether \acl{bbh} host galaxies trace stellar mass or \acl{sfr}, constraining the \acl{dtd} and shedding light on the formation channels of \aclp{bbh}.
\end{abstract}


\section{Introduction}
\label{sec:intro} 

The \ac{lvk} Collaboration's third observing run has increased the number of compact binary mergers detected in \acp{gw} to almost one hundred \citep{LIGOScientific:2021djp}.
Robust population analyses of these events have made great strides in understanding various aspects of the overall population, including the black hole mass function, merger rate, spin population, and evolution with redshift \citep{2023PhRvX..13a1048A}.
However, the formation channels for these mergers remain poorly understood.
Plausible theories include isolated evolution of main sequence stars in binaries (e.g., \citet{2016MNRAS.460.3545D, 2016ApJ...824L..10W, 2016Natur.534..512B, 2017NatCo...814906S}), dynamical exchanges in dense star clusters (e.g., \citet{2016ApJ...824L...8R, 2017MNRAS.464L..36A, 2019MNRAS.487.2947D, 2021NatAs...5..749G}), black hole encounters in gaseous environments such as \ac{agn} (e.g., \citet{2017ApJ...835..165B, 2019ApJ...878...85S, 2019PhRvL.123r1101Y, 2020ApJ...898...25T}), among others (see \citet{2021hgwa.bookE..16M} for a thorough review).

Associating a merger with its host galaxy can help build a much more complete picture of its natal state and evolution, aiding our understanding of how these compact binaries were formed.
Only one \ac{gw} event, the multi-messenger \ac{bns} merger GW170817, has been associated with its host galaxy \citep{LIGOScientific:2017vwq, GBM:2017lvd}.
While \ac{bns} mergers like GW170817 are expected to produce \ac{em} waves along with \acp{gw}, allowing for relatively straightforward host identification, all \ac{bbh} and most \ac{nsbh} mergers are likely to be \ac{em}-dark \citep{foucart2020brief, Biscoveanu:2022iue, Chen:2021fro, Fragione:2021cvv}.
With current-generation \ac{gw} detectors, pinning down a host galaxy for a merger without an \ac{em} counterpart is nonviable: at design sensitivity with a four-detector network, the best localized \acp{bbh} will have 90\% localization volumes of $\mathcal{O}(10^3)$ Mpc$^{3}$ \citep{Petrov:2021bqm}, insufficient to associate a merger with its host.

\Ac{xg} \ac{gw} detectors such as \ac{ce} \citep{Evans:2021gyd} and \ac{et} \citep{Punturo:2010zz} promise to unveil almost every \ac{bbh} merger out to redshifts greater than 10 \citep{Vitale:2016icu, Regimbau:2016ike,2021CQGra..38e5010C}. 
This data will be immensely powerful for population studies, but there will also be events at low redshifts which are detected at such high \acp{snr} that they will be localized to miniscule volumes, allowing for identification of the host galaxy with \acp{gw} alone \citep{Chen:2016tys, Vitale:2018nif, Borhanian:2020vyr}. 
We show in this work that these \ac{xg} detectors will prove to be powerful engines for finding host galaxies of \acp{cbc} such as \ac{bbh} mergers.

The host of a \ac{cbc} provides rich information about how the progenitors came to be and how they evolved.
For example, some theories predict that the dense, gaseous environments of \ac{agn} disks can catalyze the formation and inspiral of \acp{bbh} \citep{2017ApJ...835..165B, 2019ApJ...878...85S, 2019PhRvL.123r1101Y}.
In some cases, such a \ac{bbh} merger may produce an \ac{em} counterpart \citep{2019ApJ...884L..50M, 2020PhRvL.124y1102G}.
However, it can be difficult to disentangle such a counterpart with usual \ac{agn} variability or flares, making it challenging to verify these predictions \citep{2021ApJ...914L..34P}.
\Ac{gw}-only localizations of \ac{bbh} mergers to \ac{agn} hosts would provide strong evidence for the existence of this channel.

Even for individual hosts not as unique as \ac{agn}, building up demographics of hosts can be powerful, providing insight into characteristics such as late-type vs early-type galaxies, host stellar mass, and \ac{sfr}.
One useful parameterization connecting these host observables to binary formation and evolution is the \ac{dtd}, which describes the time between the zero-age main sequence formation of the stars in the binary and their eventual merger as compact objects.

The connection between host observables and the \ac{dtd} can be seen in e.g., the \ac{sfr} of hosts.
If most hosts are found to have low \ac{sfr}, the \ac{dtd} should have more support for longer delay times, since the star formation which produced the progenitor stars could have been quenched by the time the binary's merger occurs.
On the other hand, star forming hosts lend support to shorter delay times, since the host galaxy will not have evolved substantially between the formation of the individual stars in the binary and their merger.

Since the lifetime of the massive stars that result in \ac{gw}-detectable binaries is short ($\lesssim$ 10~Myr), the \ac{dtd} of \acp{cbc} essentially depends on the separation of the binary at the start of the inspiral, and the duration of the subsequent inspiral.
The \ac{dtd} is typically parameterized by a power law, $dN/dt \propto t^\alpha$, where $N$ is the number of mergers, $t$ is the time, and $\alpha$ is the power law index.
This parameterization is based on the distribution of initial separations of massive binaries and inspiral physics from general relativity, and has also been confirmed by simulations \citep{PhysRev.136.B1224, 1992ApJ...389L..45P, 2022ApJ...940L..18Z}.
A minimum delay time $t_{\rm min}$ is often used to account for the lifetime of the massive progenitors and the time required to form the resulting compact object binary \citep{Safarzadeh:2019znp, 2022ApJ...940L..18Z}.

A number of studies have found that the characterization of \ac{cbc} hosts can lead to robust constraints on the \ac{dtd}.
\citet{Adhikari:2020wpn} show from galaxy simulations that $\mathcal{O}(100)$ events with hosts will constrain the \ac{dtd}, parameterized as $dN/dt \propto (t - t_{\rm min})^\alpha$, to Gyr precision in the minimum time delay $t_{\rm min}$.
\citet{McCarthy:2020jwq} predict a similar constraint using an observed galaxy catalog, finding that $\mathcal{O}(100)$ events will constrain $\alpha$ and $t_{\rm min}$ to 10\% precision.
Using a combination of population synthesis codes and simulated galaxy catalogs, \citet{Artale:2019doq, Artale:2019tfl} connect the \ac{sfr} of a given host galaxy to a merger's time delay, finding that the \ac{dtd} can vary greatly depending on the classification of the host galaxy.
Other works such as those from \citet{Lamberts:2016txh}, \citet{Cao:2017ztl}, \citet{Safarzadeh:2019pis, Safarzadeh:2019zif, Safarzadeh:2019znp}, and \citet{Fishbach:2021mhp} have explored various aspects of hosts of binary mergers, including constraints on the \ac{dtd} and hosts of \acp{bbh}.
Evaluating these predictions with observations of mergers and their hosts will be a major component of \ac{xg} science.

Finding \ac{cbc} hosts with known redshifts is also useful for cosmology, particularly in measuring the Hubble constant $H_0$. 
Most cosmological studies with \acp{bbh} employ either a ``spectral siren'' method by fitting the black hole mass function to the \ac{bbh} population \citep{Farr:2019twy, You:2020wju,Ezquiaga:2022zkx} or a ``dark siren'' method by using galaxy catalogs and the \ac{gw} localization \citep{Oguri:2016dgk, 2019ApJ...871L..13F, DES:2019ccw, 2021ApJ...909..218A, 2024MNRAS.528.3249A}.
By uniquely identifying host galaxies using \ac{xg} detectors, \acp{bbh} can be used as ``golden dark sirens'' to directly measure $H_0$ \citep{2017PhRvD..96j1303N, 2023arXiv230710421G}.

Previous studies have explored the localization capabilities of \ac{xg} detectors and shown their promise.
\citet{Vitale:2018nif} showed with full Bayesian parameter estimation that certain events, such as the ``gold-plated'' GW150914, can be localized to within 1\% uncertainty in distance and to a 90\% sky area smaller than $10^{-2}$ deg$^2$ for a network comprised of \ac{ce} and \ac{et}.
Using Fisher matrix analyses, \citet{Borhanian:2022czq}, \citet{2022NatSR..1217940P}, and \citet{2023arXiv230710421G} show that various \ac{xg} detector networks will be able to localize hundreds to thousands of \acp{bbh} per year to within 1\% in distance and 0.1 deg$^2$ on the sky.
In particular, \citet{Borhanian:2022czq} and \citet{2023MNRAS.524.3537G} point out that \ac{xg} localizations of \acp{bbh} and \acp{nsbh} respectively could be powerful tools for resolving the Hubble tension (see \citet{2021CQGra..38o3001D} for a review).
\citet{2022ApJ...940...17C} investigate the feasibility of identifying hosts of lensed \acp{bbh} using \ac{xg} \ac{gw} detectors and next-generation galaxy surveys, finding that enough lensed \acp{bbh} and hosts may be detected to begin to distinguish \ac{bbh} formation channels. 

These highly precise \ac{gw} localizations constrain the possible site of the merger to a small volume in space, where for some events, only a single possible host galaxy will exist.
This volume can then be cross-matched with existing galaxy catalogs to identify and characterize the host galaxy.
As new \ac{em} facilities such as the Vera Rubin Observatory come online, galaxy catalogs will continue to improve \citep{2019ApJ...873..111I}.
Where catalogs are not complete, targeted \ac{em} observations could search for and characterize the host, and reveal its morphology, stellar mass $M_*$, \ac{sfr}, color, velocity dispersion, and local density around the host \citep{Adhikari:2020wpn}.

In this paper, we use a Bayesian analysis to determine the 3-dimensional localization volumes of \acp{nsbh} and  \acp{bbh} in the \ac{xg} era.
From these localizations, we estimate the rate of mergers that can be directly associated with their hosts, using population estimates put forth by \citet{2023PhRvX..13a1048A} and \citet{Biscoveanu:2022iue}.
We find that a \ac{bbh} merger will be localized to a volume smaller than 100 Mpc$^3$, the average volume for one potential host galaxy, every eight days, and an \ac{nsbh} merger every 1.5 months.
In Sec.~\ref{sec:methods} we describe the simulation and localization procedure.
Then, we present the results and show how they can be used to constrain population synthesis models in Sec.~\ref{sec:results}.
We discuss the implications and limitations of our simulation and conclude in Sec.~\ref{sec:discussion}.

\section{Methods} 
\label{sec:methods}

In our simulations, we assume one \ac{ce} placed at the current LIGO Livingston site and one \ac{et} at the current Virgo site, using the \ac{ce} and ET-D sensitivity curves from \citet{LIGOScientific:2016wof}.
To simulate the \acp{bbh}, we use the population models released by \citet{2023PhRvX..13a1048A}.
We choose the maximum likelihood distribution from the binary masses and redshift for the \texttt{Powerlaw+Peak} model; the mass model and power-law redshift model are described in \citet{2023PhRvX..13a1048A} in Appendix B.1.b and Eq.~8 respectively.
We list the chosen parameters for these distributions in Table~\ref{tab:pp_params}.
We simulate 1500 \acp{bbh} events by sampling the primary mass $m_1$, mass ratio $q$, and redshift $z$ from this distribution.

\begin{table*}
    \centering
    \caption{Parameters of the maximum likelihood distribution for the \texttt{Powerlaw+Peak} model used for drawing \acp{bbh} in the simulation, with descriptions from \cite{2023PhRvX..13a1048A}.}
    \label{tab:pp_params}
    \begin{tabular}{clc}
    \hline
    \hline
    Parameter & Description & Value \\
    \hline
    $\alpha$ & Power-law index for primary mass distribution & 3.55 \\
    $\beta$ & Power-law index for mass ratio distribution & 0.76 \\
    $m_{\rm min}$ & Minimum mass in power-law component [M$_\odot$] & 4.82 \\
    $m_{\rm max}$ & Maximum mass in power-law component [M$_\odot$] & 83.1 \\
    $\lambda_{\rm peak}$ & Fraction of systems in Gaussian component & 0.019 \\
    $\mu_{\rm m}$ & Mean of the Gaussian component in the primary mass distribution [M$_\odot$] & 34.5 \\
    $\sigma_{\rm m}$ & Width of the Gaussian component in the primary mass distribution [M$_\odot$] & 1.8\\
    $\delta_m$ & Range of mass tapering at the lower end of the mass distribution & 5.45 \\
    $\kappa$ & Power law index for redshift evolution & 3.43 \\
    $\mathcal{R}_0$ & Merger rate density at redshift zero [Gpc$^{-3}$ yr$^{-1}$] & 18.2 \\
    \hline
    \hline
    \end{tabular}
\end{table*}

For the \ac{nsbh} simulation, we use the population model defined in \citet{Biscoveanu:2022iue} to simulate 1500 events.
\citeauthor{Biscoveanu:2022iue} use a truncated power-law (parameterized by power-law index $\alpha$, minimum BH mass $m_{\rm BH, min}$, and maximum BH mass $m_{\rm BH, max}$, as described in Eq.~3 of their paper) to describe the BH mass distribution, and a truncated Gaussian model (described by mean $\mu$, standard deviation $\sigma$, the maximum neutron star mass $m_{\rm{NS, max}}$ and the drawn black hole mass $m_{\rm BH}$, in Eq.~4 of their paper) for the pairing function between the black hole and the neutron star.
The redshift distribution is assumed to be uniform in comoving volume and source frame time.
This is an appropriate assumption since the \ac{nsbh} rate evolution is not expected to be significant within the relevant redshift range. 
Similar to the \ac{bbh} simulation, we use the population hyperparameters from the maximum likelihood distribution to sample $m_1$, $q$, and $z$. 
The rate of \ac{nsbh} mergers is taken from the median of the posterior distribution.
These parameters are described in Table~\ref{tab:nsbh_params}.

\begin{table*}
    \centering
    \caption{Parameters of the chosen distribution for the \texttt{Gaussian Mass Ratio} model from \citet{Biscoveanu:2022iue} used for drawing \acp{nsbh} in the simulation. The merger rate density is taken from the median of the posterior rate distribution; all other parameters are from the maximum likelihood distribution.}
    \label{tab:nsbh_params}
    \begin{tabular}{clc}
    \hline
    \hline
    Parameter & Description & Value \\
    \hline
    $\alpha$ & Power-law index for primary mass distribution & -0.247 \\
    $m_{\rm min}$ & Minimum black hole mass [M$_\odot$] & 5.89 \\
    $m_{\rm max}$ & Maximum black hole mass [M$_\odot$] & 9.80 \\
    $m_{2,\rm max}$ & Maximum neutron star mass [M$_\odot$] & 1.98 \\
    $\mu$ & Mass ratio mean & 0.430 \\
    $\sigma$ & Mass ratio standard deviation & 0.530\\
    $\mathcal{R}_0$ & Merger rate density at redshift zero [Gpc$^{-3}$ yr$^{-1}$] & 37.29 \\
    \hline
    \hline
    \end{tabular}
\end{table*}

For both types of events, we use the \texttt{IMRPhenomD} waveform \citep{Husa:2015iqa, Khan:2015jqa} to simulate the \ac{gw} signals.
The arrival time of the events are sampled uniformly in one year, the right ascensions and declinations are sampled uniformly across the sky, the polarization angle $\psi$ is uniform between 0 and $2\pi$, and the inclination angle $\iota$ is uniform in $\cos(\iota)$.
We neglect the effect of spins to reduce computational costs.
The use of information from spins has been shown to improve localizations \citep{2008ApJ...688L..61V, 2018ApJ...854L..25P}, meaning that a future study involving spins could improve our results.
Since events at large distances will have \acp{snr} too low to be localized to the volume of a single galaxy, we limit our simulation to $z = 0.5$ for \ac{bbh} and $z=0.3$ for \ac{nsbh}.
(This choice is validated by our results, which show that almost all events with single-host localizations are located within 1~Gpc, far below $z = 0.3$.)
We assume \texttt{Planck18} cosmology \citep{Planck:2018vyg} in the simulations.

These 1500 \ac{bbh} events within $z = 0.5$ correspond to 1.25 years of observations using the merger rate density at redshift zero $\mathcal{R}_0$ from the maximum likelihood, listed in Table~\ref{tab:pp_params}.
For the \acp{nsbh}, the 1500 events within $z=0.3$ will be observed in 6.26 years of \ac{xg} detector operation.

We then produce the three-dimensional localization uncertainties of these events using the localization algorithm first presented in \citet{Chen:2015nlv} and \citet{Chen:2016tys}.
The algorithm is a maximum likelihood method which takes as inputs the detector-frame masses, detector arrival time differences, phase differences, and \acp{snr} at each detector,
then computes a three-dimensional sky localization.
While it is not a full GW parameter estimation tool (e.g., including the mass and spin estimate) like LALInference \citep{Veitch:2014wba} or Bilby \citep{Ashton:2018jfp, Romero-Shaw:2020owr}, the localization algorithm only takes $\mathcal{O}(1-10)$ minutes per event, even for \ac{xg} \acp{snr} up to $\sim 1000$, and gives localization precisions consistent with those from full parameter estimation, as shown in \citet{Chen:2015nlv}.
In addition, this method results in realistic three-dimensional localizations, where the localization regions are comprised of ``voxels'' which have sizes that increase with distance.
This is in contrast to Fisher Information studies which assume Gaussian distributions in distance.

\section{Results} 
\label{sec:results}

Understanding the rate of \ac{xg} host galaxy identification requires an estimate of the average volume in which only a single potential \ac{bbh} or \ac{nsbh} host resides.
Following \citet{Gehrels:2015uga} and \citet{Chen:2016tys}, we integrate the Schechter luminosity function \citep{Schechter:1976iz} describing the density of galaxies of a given luminosity $L$, written as $\rho(x)dx = \phi^* x^a e^{-x} dx$, where $x = L/L^*$ and $L^*$ is a characteristic luminosity describing the change between the exponential and power-law behaviors of $\rho$.
We use the same $\phi^*$, $a$, and $L^*$ values from $B$-band observations of nearby galaxies as in \citet{Gehrels:2015uga}.
Integrating the luminosity function, we find that 86\% of the total galaxy luminosity is contained in galaxies with $L > 0.12 L^*$.
Then, integrating the Schechter function to find the average number density of $L > 0.12 L^*$ galaxies, we arrive at $\phi^* \Gamma(a+1, 0.12)$ = 0.01 Mpc$^{-3}$ (where $\Gamma$ is the incomplete gamma function, see \citet{Gehrels:2015uga} for further details).
Since luminosity is robustly correlated with stellar mass, we make the assumption that almost all \ac{bbh} and \ac{nsbh} mergers occur in galaxies with $L > 0.12 L^*$.
Thus, we assume 100 Mpc$^3$ is the average volume that contains only one potential host galaxy.

Of the 1500 simulated \acp{bbh} in this study, 56 had 90\% probability localization comoving volumes smaller than 100 Mpc$^3$.
26 \ac{bbh} events were localized to 90\% volumes within 10 Mpc$^3$, with the best localized \ac{bbh} event having a 90\% volume of 0.015 Mpc$^3$.
This event was at a distance of 353 Mpc and was recovered with a network \ac{snr} of 923.
The farthest \ac{bbh} merger localized to within 100~Mpc$^3$ had a luminosity distance of 2464~Mpc, corresponding to a redshift of 0.43, showing the utility of this method beyond just the local universe.
48 of the 1500 simulated \acp{nsbh} were localized to within the single-host volume of 100~Mpc$^3$.
20 of these events were localized to within 10 Mpc$^3$.
The closest \ac{nsbh} merger, at a luminosity distance of 63.5 Mpc, was localized to a volume of $3.5 \times 10^{-3}$~Mpc$^3$; the farthest \ac{nsbh} localized to within 100~Mpc$^3$ was at a luminosity distance of 634 Mpc.

\begin{figure}
    \centering
    \includegraphics[width=\linewidth]{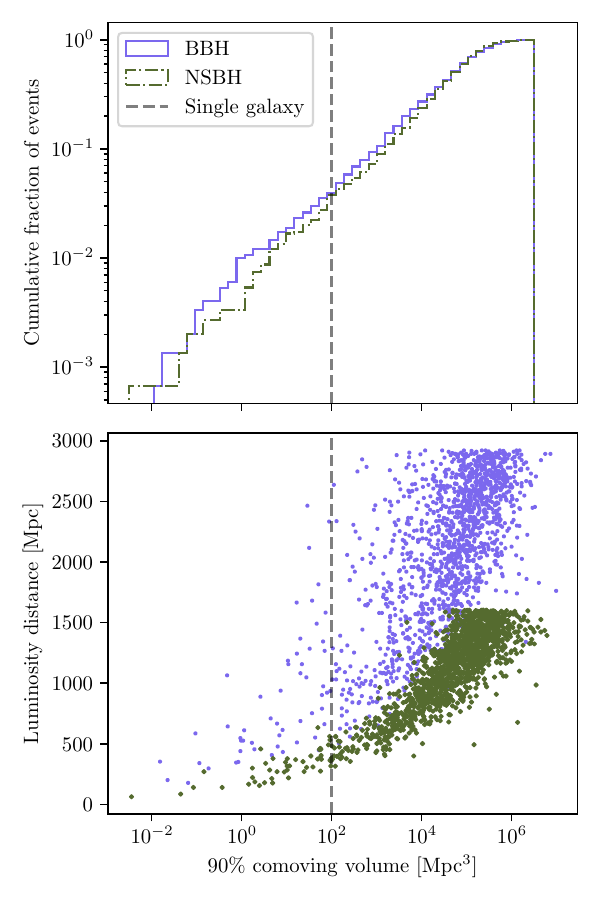}
    \caption{\textit{Top: } Cumulative distribution of 90\% probability localization comoving volumes for \acp{bbh} distributed uniformly in comoving volume within $z = 0.5$ and \acp{nsbh} within $z=0.3$, as localized by a network of \ac{ce} and \ac{et}.
    The dashed line is at 100 Mpc$^{3}$, the fiducial volume in which only a single potential host galaxy is contained.
    Approximately 4\% of \ac{bbh} events within $z = 0.5$ and 3\% of \ac{nsbh} events within $z = 0.3$ are localized to within that volume.
    \textit{Bottom: } 90\% comoving volumes versus the event's luminosity distance for \ac{bbh} and \ac{nsbh}.
    }
    \label{fig:volume}
\end{figure}

\begin{figure}
    \centering
    \includegraphics[width=\linewidth]{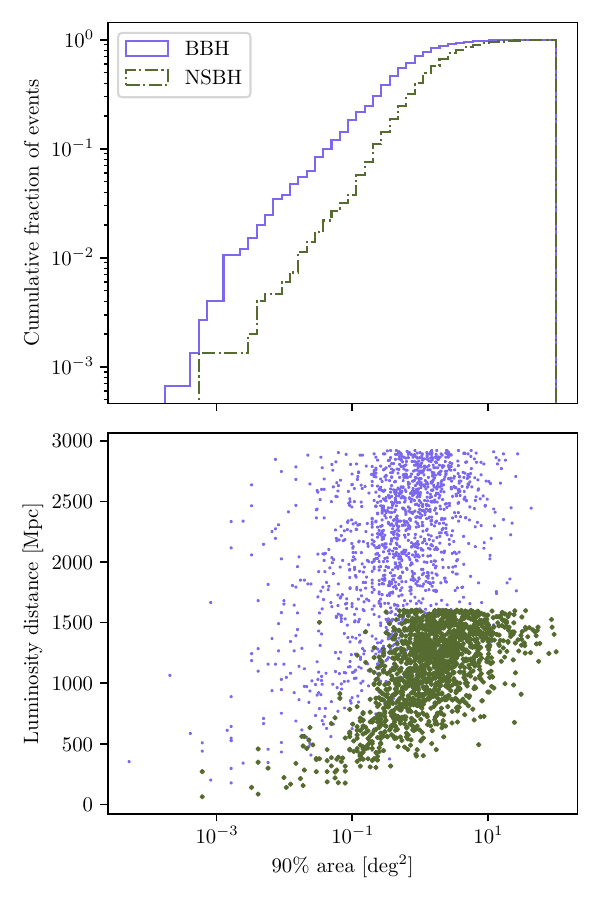}
    \caption{\textit{Top: } Cumulative distribution of 90\% probability localization areas for \acp{bbh} within $z = 0.5$ and \acp{nsbh} within $z=0.3$, as localized by a network of \ac{ce} and \ac{et}.
    \textit{Bottom: } 90\% localization areas versus the event's luminosity distance for \ac{bbh} and \ac{nsbh}.
    }
    \label{fig:area}
\end{figure}

For both \acp{bbh} and \acp{nsbh}, the distribution of 90\% volumes is shown in Fig.~\ref{fig:volume} and the distribution of 90\% areas is shown in Fig.~\ref{fig:area}.
These figures also show the trend of increasing area and volume with luminosity distance, as expected.

As mentioned above, these 1500 \ac{bbh} events (within $z = 0.5$) will be detected in 1.25 years of \ac{xg} observations.
With 56 out of 1500 (3.7\%) of such events being localized to within a single host galaxy, an average of 45 single-host \ac{bbh} events will be detected every year; this is once every eight days.
48 of the 1500 (3.2\%) of the \acp{nsbh} within $z = 0.3$ were localized to within 100 Mpc$^{-3}$ for a rate of 7.7 per year, or once every month and a half.

Studies of host galaxies and their relationship with compact binary mergers by \citet{Adhikari:2020wpn}, \citet{McCarthy:2020jwq}, and \citet{Safarzadeh:2019znp}, as mentioned in Sec.~\ref{sec:intro}, give estimates for the number of mergers with hosts required to constrain the \ac{dtd}.
Though these studies focus on \ac{bns} due to the comparative ease of finding hosts using \ac{em} counterparts, their methods and results are extendable to \ac{em}-dark \ac{bbh} and \ac{nsbh} mergers.

\citet{Adhikari:2020wpn} and \citet{McCarthy:2020jwq} both predict that with only 10 events, which we find can be observed in about two and a half months for \acp{bbh} and just over a year for \acp{nsbh}, will be enough to make distinctions between formation channels which trace various galaxy properties.
In particular, \citet{Adhikari:2020wpn} find that 10 events will be enough to distinguish between hosts that are weighted by \ac{sfr} or stellar mass from a random sample of hosts.
With $\sim 30$ events (eight months for \acp{bbh} and 3.9 years for \acp{nsbh}), we will be able to discern whether hosts tend to be stellar- or halo-mass weighted.
Studies have found that halo mass correlates with globular cluster abundance \citep{2013ApJ...772...82H, 2014ApJ...787L...5H}.
A halo mass-weighted sample of hosts might then signify that globular clusters are a more likely site of \ac{bbh} or \ac{nsbh} formation compared to the galactic field \citep{Adhikari:2020wpn}.

\citet{Safarzadeh:2019znp} predict that with a fixed $t_{\rm min}$, the \ac{dtd} power law index $\alpha$ will be constrained to within 30\% with $\mathcal{O}(300)$ hosts---about seven years of observations of \acp{bbh} and 39 years of \acp{nsbh}.
\citet{McCarthy:2020jwq} and \citet{Adhikari:2020wpn} are more optimistic, finding that $\mathcal{O}(100)$ events will be enough to constrain $\alpha$ to about ten percent precision.
$t_{\rm min}$, despite being degenerate with $\alpha$, will also be constrained, though to a lesser degree.
Thus with two years of observations of \acp{bbh} and 13 years of observations of \acp{nsbh}, their respective \ac{dtd} slopes may be determined to 10\% or tighter, and the minimum delay time to within a Gyr.

\section{Discussion and conclusion} 
\label{sec:discussion}

Our results are in broad agreement with Fisher matrix calculations from \citet{Borhanian:2022czq}, \citet{2023arXiv230710421G}, and \citet{2022NatSR..1217940P}.
While the above studies do not directly compute 3D volume localizations, agreement in our area localizations lends support to the robustness of the 3D localization results presented in this paper.

Beyond identifying the host galaxy with \acp{gw} alone, it is also possible to determine the offset between the merger site and the center of its host galaxy with precise localizations, as can be done for events with \ac{em} counterparts \citep{LIGOScientific:2017apx}, and which have been useful in studies of short gamma-ray bursts \citep{2002AJ....123.1111B, 2013ApJ...776...18F, 2020ApJ...904..190Z}.
The best localized \ac{bbh} event in our sample has a 90\% localization area of $5.1 \times 10^{-5}$ deg$^{2}$ at a distance of 353~Mpc.
Approximating this as a square and taking the root, this area corresponds to an angular resolution of 26 arcseconds.
At 353~Mpc, this can resolve an offset of 43~kpc from the center of a potential host galaxy.

In addition to formation channels, constraining cosmology is another incentive for identifying hosts of \acp{bbh} and \acp{nsbh}.
With our results showing that the identification of a unique host galaxy using \acp{gw} alone will be commonplace with \ac{xg} detectors, we can use these \acp{bbh} and \acp{nsbh} as standard sirens \citep{Schutz:1986gp, Holz:2005df, 2017Natur.551...85A, Chen:2017rfc}, since a precise redshift can be directly measured from the uniquely identified host galaxy.
Cosmological measurements made from typically \ac{em}-dark \ac{bbh} and \ac{nsbh} mergers can then be used as a consistency check against those computed from \ac{bns} standard sirens, and help to mitigate any possible systematic errors in a given method. 
For example, methods relying on the identification of an \ac{em} counterpart may suffer from biases due to the inclination angle \citep{2020PhRvL.125t1301C, 2023arXiv230710402C}, which will have no impact on events with \ac{gw}-only localizations.

One potential extension of our study involves the use of \ac{gw} waveform models which include higher-order modes, especially for \acp{bbh} and \acp{nsbh} with highly asymmetric masses.
While these models are more expensive to compute, they have been shown to tighten parameter estimation constraints on localization \citep{2009PhRvD..79h4032A, 2020PhRvD.101j3004K, Borhanian:2020vyr}.

In simulating our \ac{bbh} and \ac{nsbh} populations, we necessarily assume a merger rate.
While the \ac{lvk} have detected enough \ac{bbh} to start placing meaningful constraints on their population distribution, the rate of \ac{nsbh} coalescences remains highly uncertain. 
We expect upcoming \ac{lvk} observing runs to significantly reduce this uncertainty.
Any future rate estimate will act as a scaling factor on the overall rate of detections; the 3\% proportion of \ac{nsbh} which will be localized to a single host galaxy for events within $z = 0.3$ remains valid as long as the inferred \ac{nsbh} population is similar to the one simulated here.

Our study neglects data analysis issues that have yet to be solved for \ac{xg} detectors.
These include the identification and isolation of individual events from a background of overlapping signals \citep{Janquart:2022nyz, Samajdar:2021egv}, \ac{gw} waveform accuracy and systematics at very high \ac{snr} \citep{ Purrer:2019jcp, Huang:2022rdg, Hu:2022rjq}, and others mentioned in \citet{Couvares:2021ajn}. 
Our analyses focus primarily on the localization of a particular set of \ac{gw} events which tend to be at low redshift and recovered with high \ac{snr}. 
For these events, some as-yet unsolved \ac{xg} analysis problems will not have significant effects; for example, they will stand out from any overlapping sources or noise.
Additionally, we expect that many of these challenges will be solved by the time that \ac{xg} detectors begin observations in the mid-2030s.

We note that our estimate of the host galaxy number density stems from integrating the Schechter function down to 0.12 $L^*$.
If less luminous galaxies such as dwarf galaxies have nontrivial contributions to the compact binary population (e.g., \cite{Conselice:2019wfn}), then the average volume in which only a single host galaxy exists will be smaller than fiducial value of 100 Mpc$^{3}$.
In addition, this value is a statement of the \textit{average} number density; in reality, galaxies tend to be clustered.

For \ac{gw} localizations where galaxy catalogs are sparse, a minimal amount of telescope time will be required to characterize a potential host. 
As an example, a robust spectrum of a Milky Way-like galaxy at 2~Gpc is obtainable with an 1800 s exposure on an 8~m class telescope.

Beyond observations of \ac{cbc} hosts, \Ac{dtd} constraints can also be obtained in other ways.
Using synthetic \ac{xg} detections of \ac{bbh} mergers, \citet{2019ApJ...886L...1V} simultaneously fit the \ac{sfr} and the \ac{dtd}.
They indicated a similar or longer observation time with \ac{xg} detectors compared to our two years (for \ac{bbh} observations) to reach a similar 10\% constraint on a characteristic \ac{dtd} delay time.
Since the two methods rely on different assumptions, it will be useful to compare the two approaches once \ac{xg} data begins to flow to verify their results.

We note that precise localizations heavily depend on the size of the \ac{xg} detector network.
If only one \ac{xg} detector operates, the prospects for constraining \acp{bbh} and \acp{nsbh} to within a single host galaxy with \acp{gw} alone become much more bleak. 
As an example, if the current best-localized \ac{bbh} in the simulation (at 353 Mpc, localized to 0.015 Mpc$^{3}$ with the \ac{et} + \ac{ce} network) were localized with just \ac{et}, its 90\% localization volume would balloon to 19 000 Mpc$^{3}$. 
This emphasizes the importance of a global network of detectors in enabling future \ac{gw} science.

We have shown that with \ac{xg} detectors, singling out a unique galaxy as the host to a \ac{bbh} merger with \acp{gw} alone will be a regular occurrence.
Every eight days on average, an \ac{xg} network comprised of \ac{ce} and \ac{et} will detect and localize a \ac{bbh} merger to a comoving volume smaller than 100 Mpc$^{3}$, the average volume in which only a single probable host galaxy resides.
This information will be invaluable in constraining the \ac{dtd} and in determining whether hosts of these mergers trace \ac{sfr}, stellar mass, galaxy morphology, or other parameters.
Within eight months of observations, we will be able to determine if \ac{bbh} host galaxies are stellar- or halo-mass weighted for \acp{bbh}, and thus whether mergers are more likely to be formed in globular clusters versus in the fields of galaxies.
In two years of observations, the \ac{bbh} \ac{dtd} parameters may be constrained to within 30\%.
For \acp{nsbh}, for which the detection rates are much lower, a single-host localization will happen every six weeks.
Finally, this method will allow for \ac{em}-determined redshifts from hosts of \ac{em}-dark \acp{bbh} and \acp{nsbh}, proving useful for standard siren cosmology as well.

\begin{acknowledgments}
We are thankful for helpful discussions with Sylvia Biscoveanu and Colm Talbot.
G.~M. acknowledges the support of the National Science Foundation and the LIGO Laboratory.
LIGO was constructed by the California Institute of Technology and
Massachusetts Institute of Technology with funding from the National
Science Foundation and operates under cooperative agreement PHY-0757058.
HYC is supported by the National Science Foundation under Grant PHY-2308752.
The authors are grateful for computational resources provided by the LIGO Lab and supported by NSF Grants PHY-0757058 and PHY-0823459. 
Some of the results in this paper have been derived using the \texttt{healpy} \citep{2019JOSS....4.1298Z} and
\texttt{HEALPix} \citep{2005ApJ...622..759G} packages.
\end{acknowledgments}

\facilities{Cosmic Explorer, Einstein Telescope}

\software{\texttt{astropy} \citep{astropy:2013},
        \texttt{bilby} \citep{Ashton:2018jfp, Romero-Shaw:2020owr},
        \texttt{gwpopulation} \citep{Talbot:2019okv},
        \texttt{healpy} \citep{2019JOSS....4.1298Z},
         \texttt{HEALPix} \citep{2005ApJ...622..759G},
        \texttt{ligo.skymap} \citep{Singer:2015ema, Singer:2016eax, Singer:2016erz},
         \texttt{matplotlib} \citep{Hunter:2007},
         \texttt{numpy} \citep{harris2020array},
         \texttt{pandas} \citep{reback2020pandas, mckinney2010data},
         \texttt{scipy} \citep{2020SciPy-NMeth},
        }

\bibliography{singlehost}{}
\bibliographystyle{aasjournal}

\end{document}